\documentclass[11pt]{article} 

\usepackage[utf8]{inputenc} 
\usepackage{amsmath}

\usepackage{geometry} 
\usepackage{graphicx} 

\usepackage{booktabs} 
\usepackage{array} 
\usepackage{paralist} 
\usepackage{verbatim} 
\usepackage{subfig} 

\usepackage{fancyhdr} 
\usepackage{sectsty}
\allsectionsfont{\sffamily\mdseries\upshape} 

\usepackage[nottoc,notlof,notlot]{tocbibind} 
\usepackage[titles,subfigure]{tocloft} 

\title{Excess-noise suppression for a squeezed state propagating through random amplifying media via wave-front shaping}

\author{Dong Li$^{1,2}$, Song Sun$^{1,2}$, Yao Yao$^{1,2\dagger}$}

\date{$^1$Microsystems and Terahertz Research Center, China Academy of Engineering Physics,
Chengdu Sichuan 610200, P. R. China\\%
$^2$Institute of Electronic Engineering, China Academy of Engineering Physics, Mianyang Sichuan 621999, P. R. China\\
$^\dagger$yaoyao\_mtrc@caep.cn\\
\today}

\begin{document}
\maketitle

\begin{abstract}
After propagating through a random amplifying medium, a squeezed state commonly shows excess noise above the shot-noise level. Since large noise can significantly reduce the signal-to-noise ratio, it is detrimental for precision measurement. To circumvent this problem, we propose a noise-reduction scheme using wavefront shaping. It is demonstrated that the average output quantum noise can be effectively suppressed even beyond the shot-noise limit. Both the decrease on amplification strength and the increase on input squeezing strength can give rise to a decrease in the suppressed average quantum noise. Our results not only show the feasibility of manipulating the output quantum noise of random amplifying media, but also indicate potential applications in quantum information processing in complex environments, such as, quantum imaging, quantum communication, and quantum key distribution.
\end{abstract}

\section{Introduction}
The random medium exhibits unusual transmission properties which couples light into different channels randomly by multiple scattering. Previously, light scattering was considered harmful, since it may distort the incident wavefront and result in a speckle pattern. Later, it is shown that light scattering could also play a positive role in many applications. For instance, (1) in imaging, it can improve the resolution by overcoming the traditional diffraction limit, owing to the increased effective aperture number by multiple scattering \cite{van2011,park2014}; (2) in optical communication, it provides the possibility to increase the capacity by the raising number of scattered modes that carry the information \cite{simon2001,two2002}. In addition, light scattering can also be applied in other fields, such as, secure authentication \cite{goorden2014,yao2016}, high-speed random number generator \cite{argyris2010,xiang2015}, programmable optical circuit \cite{huisman2015,marcucci2020}. Therefore, light transport through random media has become an active subject from both theoretical and experimental perspectives.

In particular, the random amplifying media (RAMs) have attracted considerable attention because nonlinearity or amplification provides an additional degree of freedom for coherent control of mesoscopic transport \cite{renthlei2015,liew2015}. By adjusting the amplification strength, one could conveniently manipulate the transmission properties of light which could benefit for many potential applications, such as, random laser \cite{cao2003,luan2015,churkin2015}.

Recently, coherent-state light propagation through a RAM has been explored from different aspects. For example, Liew \textit{et al.} \cite{liew2015} investigated the effect of amplification on the reflection properties. It was revealed that the amplification could minimize the reflectance of the random medium by destructive interference. Burkov \textit{et al.} \cite{burkov1997} studied the correlation of scattered light. It was demonstrated that the angular correlation has a power-law decay and exhibits oscillations. Patra \textit{et al.} \cite{patra1999} analyzed the quantum noise of the scattered light. It is found that the output shows excess noise related to the transmission and reflection matrices of the medium for a coherent-state input.

As a typical nonclassical state, the squeezed state is of importance since it possesses lower quantum noise in one quadrature component than that of the coherent state (\textit{or} equivalently the shot noise) \cite{walls1983,walls2007,barnett2002,lvovsky2015}. Therefore, the squeezed state can enhance signal-to-noise ratio \cite{caves81,yurke1986,xiao1987precision} and has been utilized in different applications ranging from quantum imaging \cite{beskrovnyy2005,sokolov2004} to gravitational wave detection \cite{aasi2013,barsotti2018,mehmet2018}. 

\begin{figure*}[tb]
\begin{center}
\includegraphics[width=0.8\textwidth]{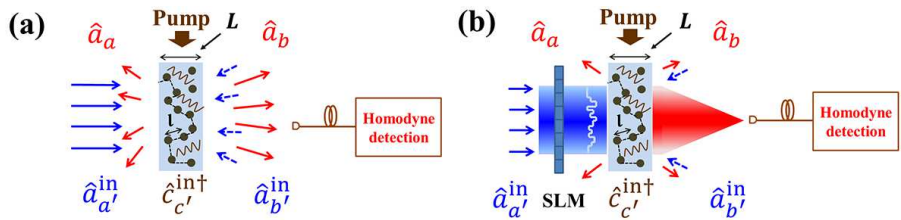}
\end{center}
\caption{Quadrature fluctuation detection of beams transmitted through a random amplifying medium (a) in the absence of WFS, (b) in the presence of WFS. $\hat{a}_{a'}^{\rm{in}}$ ($\hat{a}_{b'}^{\rm{in}}$) represent the annihilation operators of the input modes and $\hat{a}_{a}$ ($\hat{a}_{b}$) the output modes. $\hat{c}_{c'}^{\rm{in} \dagger}$ denote the creation operators of the spontaneous emission modes symbolized by the wavy-dotted lines inside the media. The random amplifying medium, with the transport mean free path $l$, the thickness $L$, the amplification length $L_a$, and the number of transmission channels $N$, is comprised by randomly distributed small active particles for light scattering and amplification. In (a), when the beams are injected, the medium amplifies and separates the light into different optical channels randomly. As a consequence, the output is in a speckle pattern. In (b), the medium amplifies and couples the beams into the desired optical paths. Hence the output presents an ordered pattern. The WFS, performed by a spatial light modulator (SLM) in (b), controls the incident phase of light. In our scheme, the focus is on the quadrature of the scattered mode, monitored by homodyne detection. }
\label{fig1}
\end{figure*}

However, the squeezed state suffers from an increase in output quantum noise after propagating through a RAM \cite{patra2000} (see Fig. \ref{fig1}(a)), which is induced by spontaneous emission and multiple scattering. It is worth pointing out that for a squeezed-state input, the input quantum noise is below the shot-noise level (SNL), whereas the output noise is always increased, even above the SNL. More interestingly, compared with the coherent state, the squeezed state initially possessing lower noise will have larger noise in the output \cite{patra2000}. As is well known, the large noise leads to a decrease in the signal-to-noise ratio which is detrimental for precision detection (e.g., high-resolution imaging). Therefore, we wonder whether there exists a method to reduce the average output noise for the squeezed-state input.

Wavefront shaping (WFS) is a promising technology for optical focusing and imaging through random media \cite{vellekoop2007,vellekoop2008,popoff2010,mosk2012,hong2017,hong2018,tzang2019,blochet2019}, which paves the way in manipulating the speckle pattern in an expected manner. Experimentally, the WFS can be performed by a spatial light modulator (SLM), as shown in Fig. \ref{fig1}(b). The SLM acts as a reconfigurable matrix of pixels to imprint desired phases on the incident wavefront. In recent decades, it has been extensively used in various optical applications, for instance, quantum simulator \cite{prl2019}, quantum data locking \cite{lum2016}, and high-resolution imaging \cite{jang2018,chen2018}. In particular, the WFS is also a common technique in the optical authentication scheme based on scattering medium \cite{goorden2014,yao2016}. 

In this work, we propose a noise-reduction scheme using WFS for the case of squeezed-state light propagating through RAMs. Comparing with Ref. \cite{patra2000}, we exploit the technique of WFS to reduce the output noise. In addition, the comparison between the linear and amplifying cases is performed. It is found that the amplifying media always have larger average quantum noise than that of the linear ones regardless of WFS. Besides, unlike the linear case where the suppressed quantum fluctuation always reaches below the SNL, the reduced quantum fluctuation can be either below or above the SNL for the amplifying case. Moreover, we provide the condition for the reduced average quantum fluctuation to reach below the SNL.

This paper is organized as follows: in Sec. 2, we briefly describe the model of propagation of quantized light through a RAM. Sec. 3 elucidates how the WFS suppresses the average quantum fluctuation of output modes. In Sec. 4, we compare the cases of amplifying and linear media. Sec. 5 is devoted to the conclusion of the main results.

\section{Theoretical model}

Fig. \ref{fig1}(a) illustrates the propagation of quantized light through a RAM. Generally, a RAM consists of randomly distributed scattering particles with amplification either in the background medium or in the particles themselves. When light propagates through a RAM, it would be multiple scattered and amplified.

To quantitatively characterize the property of a RAM, three kernel factors are introduced: the transport mean free path $l$, the thickness $L$, and the amplification length $L_a$ \cite{patra1999}. Note that different from the linear media with only two primary parameters ($L$ and $l$) \cite{xu2017a,xu2017b,li2019,li2019b}, the RAMs require an extra amplification length $L_a = \sqrt{D \tau_a}$ to account the nonlinearity \cite{patra1999}, where $1/\tau_a$ is the amplification rate and $D = c l /3$ the diffusion constant ($c$ the velocity of light in the medium).

\subsection{Propagation of quantized light through a random amplifying medium}

After propagating through a RAM, the scattered mode $b$ can be expressed as \cite{fedorov2009}
\begin{eqnarray}
\label{inout001}
\hat{a}_b = \sum_{a'=1}^{N} t_{a' b} \hat{a}_{a'}^{\rm{in}} + \sum_{b'=N+1}^{2N} r_{b' b} \hat{a}_{b'}^{\rm{in}} + \sum_{c'} v_{c' b}^{\ast} \hat{c}_{c'}^{\rm{in}\dagger},
\end{eqnarray}
where $\hat{a}_b$ indicates the annihilation operator of scattered mode $b$, $\hat{a}_{a'}^{\rm{in}}$ ($\hat{a}_{b'}^{\rm{in}}$) the annihilation operators of input modes on the left-hand side (right-) of the RAM, $\hat{c}^{\rm{in}\dagger}_{c'}$ the creation operators of spontaneous emission modes inside the RAM, $t_{a' b}$ ($r_{b' b}$) the transmission (reflection) coefficients from the input modes $a'$ ($b'$) to the output mode $b$, $v_{c' b}^{\ast}$ the connection between the spontaneous emission modes and the output mode $b$, $N$ the number of transmission channels. Noticeably, the last term on the right-hand side in Eq. (\ref{inout001}) quantifies the spontaneous emission inside the RAM, with $c'$ running over ``objects'' (e.g., atoms or molecules) and the operator $\hat{c}_{c'}^{\rm{in} \dagger}$ fulfilling the commutation relation $[\hat{c}_{i}^{\rm{in}}, \hat{c}_{j}^{\rm{in} \dagger}] = \delta_{ij}$. 

Unlike the random linear medium with only transmission and reflection coefficients ($t_{a' b}$, $r_{b' b}$), the RAM involves an additional spontaneous emission coefficient ($v_{c'b}^{\ast}$), which are subject to a constraint $\sum_{a'} |t_{a' b}|^2 + \sum_{b'} |r_{b' b}|^2 - \sum_{c'}|v^{\ast}_{c' b}|^2 = 1$ (see Appendix A). The ensemble-averaged transmission, reflection, and spontaneous-emission coefficients are given by \cite{fedorov2009}
\begin{eqnarray} 
\label{eq2a}
\overline{T_{a' b}} &=& \frac{1}{N}\frac{\sin(l/L_{a})}{\sin(L/L_a)},\\
\label{eq2b}
\overline{R_{b' b}} &=& \frac{1}{N}\frac{\sin[(L-l)/L_{a}]}{\sin(L/L_a)},\\%
\overline{V_{b}} &=& \frac{\sin(l/L_a) + \sin[(L-l)/L_a]}{\sin(L/L_a)}-1,
\label{eq2c} 
\end{eqnarray}
where $ T_{a'b}=  |t_{a'b}|^2$, $ R_{b'b}=  |r_{b'b}|^2 $, $V_b = \sum_{c'} V_{c'b}= \sum_{c'}|v_{c'b}|^2$, and the overline stands for the average over the ensemble of disorder realizations. Note that $\overline{T_{a'b}}$, $\overline{R_{b'b}}$, and $\overline{V_{c'b}}$ diverge at $L/L_{a} = \pi$ which is identify as a threshold for random-laser emission. Clearly, this analysis method based on Eq. (\ref{inout001}) can only be applied below the laser threshold (i.e., $L/L_a < \pi$). If $L/L_a$ is infinite small (ie., $L/L_a \to 0$), Eqs. (\ref{eq2a})-(\ref{eq2c}) can be rewritten as $\overline{T_{a' b}} =[1/(L/l)]/N$, $\overline{R_{b' b}} =[1-1/(L/l)]/N $, and $\overline{V_{b}} =0$, respectively, which are exactly the same as the coefficients for the linear case \cite{lodahl2006b}. Evidently, this generalized model is suitable for both the amplifying and the linear cases.

The quadrature operators are introduced as $\hat{x} = \hat{a}^{\dagger} + \hat{a}$ and $\hat{p} = i(\hat{a}^\dagger - \hat{a})$. According to Eq. (\ref{inout001}), the quadrature operators of scattered mode $b$ are then written as
\begin{eqnarray} 
\hat{x}_b=&\sum_{a'}\sqrt{T_{a'b}}[\cos\phi_{a'b}\hat{x}_{a'}^{\rm{in}}-\sin\phi_{a'b}\hat{p}_{a'}^{\rm{in}}]\nonumber\\
&+\sum_{b'}\sqrt{R_{b' b}}[\cos \phi_{b'b}\hat{x}_{b'}^{\rm{in}}-\sin\phi_{b'b}\hat{p}_{b'}^{\rm{in}}]\nonumber\\
&+\sum_{c'}\sqrt{V_{c' b}}[\cos\phi_{c'b}\hat{x}_{c'}^{\rm{in}}-\sin\phi_{c'b}\hat{p}_{c'}^{\rm{in}}],\label{xb}\\
\hat{p}_b=& \sum_{a'}\sqrt{T_{a' b}}[\sin\phi_{a' b}\hat{x}_{a'}^{\rm{in}}+\cos\phi_{a'b}\hat{p}_{a'}^{\rm{in}}]\nonumber\\
&+\sum_{b'}\sqrt{R_{b' b}}[\sin \phi_{b' b}\hat{x}_{b'}^{\rm{in}}+\cos\phi_{b'b}\hat{p}_{b'}^{\rm{in}}]\nonumber \\
&-\sum_{c'}\sqrt{V_{c'b}}[\sin\phi_{c' b}\hat{x}_{c'}^{\rm{in}}+\cos\phi_{c'b}\hat{p}_{c'}^{\rm{in}}],\label{pb}
\end{eqnarray} 
where we have defined $t_{a'b} = \sqrt{T_{a'b}} e^{i \phi_{a'b}}$, $r_{b'b} = \sqrt{R_{b'b}} e^{i \phi_{b'b}}$, $v_{c'b}^{\ast} = \sqrt{V_{c'b}} e^{-i \phi_{c'b}}$, $\hat{x}^{\rm{in}}_{c'} = \hat{c}^{\rm{in}\dagger}_{c'} + \hat{c}^{\rm{in}}_{c'}$, and $\hat{p}^{\rm{in}}_{c'} = i(\hat{c}^{\rm{in}\dagger}_{c'} - \hat{c}^{\rm{in}}_{c'})$.

\subsection{Modified propagation via wavefront shaping}

In this work we consider the situation of optical focusing through a random medium with WFS. In such a case, the expected phases, $\phi^{\rm{SLM}}_{a'}=-\phi_{a'b}$ $(a'=1,2,...,N)$ are imprinted on the incident wavefront via WFS where the output mode $b$ corresponds to the focused beam. This phase modulator exactly compensates the phase retardation in the RAM for each transmission channel which leads to a constructive interference in the output mode $b$. Correspondingly, the initial input-output relation in Eq. (1) is modified as
\begin{eqnarray}
\hat{a}_{b}^{\rm{w}} = \sum_{a'=1}^{N}{|t_{a'b }|  \hat{a}_{a'}^{\rm{in}}} + \sum_{b'=N+1}^{2N}{r_{b'b }  \hat{a}_{b'}^{\rm{in}}}+ \sum_{c'} v_{c' b}^{\ast} \hat{c}_{c'}^{\rm{in}\dagger},
\label{creation2}
\end{eqnarray}
where the superscript $w$ stands for WFS and $|t_{a'b}|$ takes the place of the original complex transmission coefficient $t_{a'b}$.


By taking into account the WFS, based on Eq. (\ref{creation2}), the quadrature operators of the scattered mode $b$ now becomes
\begin{eqnarray}
\label{xp1}
\hat{x}_b^{\rm{w}} &=& \sum_{a'}{\sqrt{T_{a'b}} \hat{x}_{a'}^{\rm{in}} } + \sum_{b'}{\sqrt{R_{b'b}} [\cos \phi_{b'b} \hat{x}_{b'}^{\rm{in}} - \sin \phi_{b'b} \hat{p}_{b'}^{\rm{in}}] } \nonumber \\ 
 && + \sum_{c'}{\sqrt{V_{c'b}} [\cos \phi_{c'b} \hat{x}_{c'}^{\rm{in}} - \sin \phi_{c'b} \hat{p}_{c'}^{\rm{in}}] },\\
\label{xp2}
\hat{p}_b^{\rm{w}} &=& \sum_{a'}{\sqrt{T_{a'b}} \hat{p}_{a'}^{\rm{in}} } + \sum_{b'}{\sqrt{R_{b'b}} [\cos \phi_{b'b} \hat{p}_{b'}^{\rm{in}} + \sin \phi_{b'b} \hat{x}_{b'}^{\rm{in}}] } \nonumber \\
&&+ \sum_{c'}{\sqrt{V_{c'b}} [\cos \phi_{c'b} \hat{p}_{c'}^{\rm{in}} + \sin \phi_{c'b} \hat{x}_{c'}^{\rm{in}}] }.
\end{eqnarray}

Note in passing that our scheme can be realized with a similar experimental setup as shown in Refs. \cite{qiao2017,peng2018,osnabrugge2019}. Nevertheless, those works mainly focusing on the enhanced intensity in the speckle pattern, whereas our work will concentrate on the reduced quantum noise of scattered modes. 

\section{Variance of quadrature of the scattered modes}

\begin{figure*}[tbh]
\begin{center}
\includegraphics[width=0.8\textwidth]{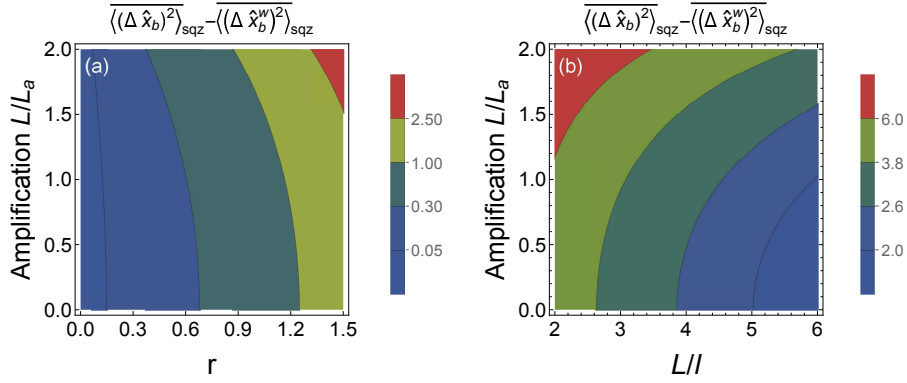} {}
\end{center}
\caption{The difference between $\overline{\langle (\Delta \hat{x}_b)^2 \rangle}$ and $\overline{\langle (\Delta \hat{x}_b^{\rm{w}})^2 \rangle}$ as a function of (a) $r$ and $L/L_a$, (b) $L/l$ and $L/L_a$, in the presence of a squeezed-state input ($|\Psi^{\rm{in}} \rangle = [\hat{D}(\alpha)\hat{S}(r)|0\rangle]^{\otimes N}$), with the number of transmission channels $N$, displacement operator $\hat{D}(\alpha) = e^{\alpha \hat{a}^{\dagger} - \alpha^{\ast} \hat{a}}$, and squeezing operator $\hat{S}(r) = e^{ (r /2)(\hat{a}^{\dagger 2} - \hat{a}^{2}) }$ (the complex number $\alpha$ being the amplitude and the real number $r$ the squeezing parameter). Parameters used are (a) $L/l=6$ and (b) $r=1.5$.}
\label{fig002}
\end{figure*}

The variance of operator $\hat{O}$ is defined as
\begin{eqnarray}
\label{var20}
\langle (\Delta \hat{O})^2 \rangle \equiv \langle \hat{O}^2 \rangle - \langle \hat{O} \rangle^2,
\end{eqnarray}
where $\hat{O} = \hat{x}_b^{\rm{w}},\hat{p}_b^{\rm{w}}$, and $\langle \hat{O} \rangle$ denotes the expectation value of $\hat{O}$. That is to say, to obtain the variances, it requires to calculate $\langle \hat{x}_b^{\rm{w}} \rangle, \langle \hat{p}_b^{\rm{w}} \rangle, \langle (\hat{x}_b^{\rm{w}})^2 \rangle$, and $\langle (\hat{p}_b^{\rm{w}})^2 \rangle$.

Assuming that the light is only injected on the left-hand side of the RAM [see Fig. \ref{fig1}(b)] and the input beams of the other side are vacuum states (i.e., $\langle \hat{x}_{b'}^{\rm{in}}\rangle = \langle \hat{p}_{b'}^{\rm{in}}\rangle=0$). According to Eqs. (\ref{xp1}) and (\ref{xp2}), the expectation values of $\hat{x}_b^{\rm{w}}$ and $\hat{p}_b^{\rm{w}}$ can be obtained
\begin{eqnarray}
\label{var2a}
\langle \hat{x}_b^{\rm{w}} \rangle &=& \sum_{a'}{\sqrt{T_{a'b}}  \langle \hat{x}_{a'}^{\rm{in}} \rangle}, \\
\label{var2b}
\langle \hat{p}_b^{\rm{w}} \rangle &=& \sum_{a'}{\sqrt{T_{a'b}}  \langle \hat{p}_{a'}^{\rm{in}} \rangle}.
\end{eqnarray}
Note that $\langle \hat{x}_b^{\rm{w}} \rangle$ ($\langle \hat{p}_b^{\rm{w}} \rangle$) is only related to the transmitted modes $\langle \hat{x}_{a'}^{\rm{in}} \rangle$ ($\langle \hat{p}_{a'}^{\rm{in}} \rangle$). 

Similarly, from Eqs. (\ref{xp1}) and (\ref{xp2}), the expectation values of $(\hat{x}_b^{\rm{w}})^2$ and $(\hat{p}_b^{\rm{w}})^2$ are found to be
\begin{eqnarray}
\label{sqzxb2}
\langle (\hat{x}_b^{\rm{w}})^2 \rangle &= & \sum_{a'a''} \sqrt{T_{a' b}T_{a'' b}} [ \langle\hat{x}_{a'}^{\rm{in}} \hat{x}_{a''}^{\rm{in}}\rangle ] \nonumber \\
&&+\sum_{b'} R_{b' b} [\cos^2 \phi_{b' b} \langle \hat{x}_{b'}^{\rm{in} 2} \rangle + \sin^2 \phi_{b' b} \langle \hat{p}_{b'}^{\rm{in} 2} \rangle \nonumber \\
&&-\cos \phi_{b' b}  \sin \phi_{b' b} \langle \hat{x}_{b'}^{\rm{in}} \hat{p}_{b'}^{\rm{in}} + \hat{p}_{b'}^{\rm{in}} \hat{x}_{b'}^{\rm{in}} \rangle] \nonumber\\
&&+\sum_{c'} V_{c' b} [\cos^2 \phi_{c' b} \langle \hat{x}_{c'}^{\rm{in} 2} \rangle + \sin^2 \phi_{c' b} \langle \hat{p}_{c'}^{\rm{in} 2} \rangle \nonumber \\ 
&&-\cos \phi_{c' b} \sin \phi_{c' b} \langle \hat{x}_{c'}^{\rm{in}}  \hat{p}_{c'}^{\rm{in}} +  \hat{p}_{c'}^{\rm{in}} \hat{x}_{c'}^{\rm{in}}\rangle ],
\end{eqnarray}
\begin{eqnarray}
\label{sqzpb2}
\langle (\hat{p}_b^{\rm{w}})^2 \rangle &= & \sum_{a'a''} \sqrt{T_{a' b}T_{a'' b}} [ \langle\hat{p}_{a'}^{\rm{in}} \hat{p}_{a''}^{\rm{in}}\rangle ] \nonumber\\
&&+\sum_{b'} R_{b' b}  [\cos^2 \phi_{b' b} \langle \hat{p}_{b'}^{\rm{in} 2} \rangle + \sin^2 \phi_{b' b} \langle \hat{x}_{b'}^{\rm{in} 2} \rangle \nonumber \\
&&+ \cos \phi_{b' b}  \sin \phi_{b' b} \langle \hat{x}_{b'}^{\rm{in}} \hat{p}_{b'}^{\rm{in}} + \hat{p}_{b'}^{\rm{in}} \hat{x}_{b'}^{\rm{in}} \rangle] \nonumber\\
&&+  \sum_{c'} V_{c' b} [\cos^2 \phi_{c' b} \langle \hat{p}_{c'}^{\rm{in} 2} \rangle + \sin^2 \phi_{c' b} \langle  \hat{x}_{c'}^{\rm{in} 2} \rangle \nonumber \\
&& + \cos \phi_{c' b} \sin \phi_{c' b} \langle \hat{x}_{c'}^{\rm{in}}  \hat{p}_{c'}^{\rm{in}} +  \hat{p}_{c'}^{\rm{in}} \hat{x}_{c'}^{\rm{in}}\rangle ],
\end{eqnarray}
which are universal for arbitrary input state.

Inserting Eqs. (\ref{var2a}) and (\ref{sqzxb2}) into Eq. (\ref{var20}) yields the variance $\langle (\Delta \hat{x}_b^{\rm{w}})^2\rangle$ as
\begin{eqnarray}
\label{sqzxb2b1}
\langle (\Delta \hat{x}_b^{\rm{w}})^2 \rangle &=& 
\sum_{a'}{T_{a'b}\langle (\Delta \hat{x}_{a'}^{\rm{in}})^2 \rangle } \nonumber\\
&&+ \sum_{a'\neq a''}  2\sqrt{T_{a'b} T_{a''b}} [  {\rm{cov}} (\hat{x}_{a'}^{\rm{in}}, \hat{x}_{a''}^{\rm{in}})  ]    \nonumber \\
&&+ \sum_{b'} R_{b'b}[\cos^2 \phi_{b'b}\langle (\Delta \hat{x}_{b'}^{\rm{in}})^2 \rangle \nonumber\\
&&+ \sin^2 \phi_{b'b} \langle (\Delta \hat{p}_{b'}^{\rm{in}})^2 \rangle  \nonumber \\
&&- 2\cos \phi_{b'b} \sin \phi_{b'b} {\rm{cov}}(\hat{x}_{b'}^{\rm{in}}, \hat{p}_{b'}^{\rm{in}}) ] \nonumber\\
&& + \sum_{c'}V_{c'b}[\cos^2 \phi_{c'b}\langle (\Delta \hat{x}_{c'}^{\rm{in}})^2 \rangle \nonumber \\
&& + \sin^2 \phi_{c'b}\langle (\Delta \hat{p}_{c'}^{\rm{in}})^2 \rangle \nonumber\\
&& - 2\cos \phi_{c'b} \sin \phi_{c'b} {\rm{cov}}(\hat{x}_{c'}^{\rm{in}}, \hat{p}_{c'}^{\rm{in}}) ],
\end{eqnarray}
where the covariance function is defined as ${\rm{cov}}(\hat{Y},\hat{Z}) \equiv \frac{1}{2} (\langle \hat{Y} \hat{Z} \rangle +\langle \hat{Z} \hat{Y} \rangle) - \langle \hat{Y}\rangle \langle \hat{Z} \rangle $. 

By averaging over the ensemble of RAMs, we obtain
\begin{eqnarray}
\label{mean002az}
\overline {\langle (\Delta \hat{x}_b^{\rm{w}})^2 \rangle}
&=& \overline{\sum_{a'} T_{a'b} \langle (\Delta \hat{x}_{a'}^{\rm{in}})^2 \rangle} \nonumber \\
&& + \overline{\sum_{a'\neq a''}  2\sqrt{T_{a'b} T_{a''b}}   {\rm{cov}} (\hat{x}_{a'}^{\rm{in}}, \hat{x}_{a''}^{\rm{in}})   }  \nonumber \\
&& + \overline{\sum_{b'} \frac{1}{2} R_{b'b} [\langle (\Delta \hat{x}_{b'}^{\rm{in}})^2 \rangle + \langle (\Delta \hat{p}_{b'}^{\rm{in}})^2 \rangle]} \nonumber\\
&& + \overline{\sum_{c'} \frac{1}{2}V_{c'b} [\langle (\Delta \hat{x}_{c'}^{\rm{in}})^2 \rangle + \langle (\Delta \hat{p}_{c'}^{\rm{in}})^2 \rangle]},
\end{eqnarray}
where we have used $\overline{\sin^2 \phi_{b'b} } = \overline{\sin^2 \phi_{c'b} } = \overline{\cos^2 \phi_{b'b} } =\overline{\cos^2 \phi_{c'b} } = 1/2$ and $\overline{\cos \phi_{b'b} \sin \phi_{b'b} }= \overline{\cos \phi_{c'b} \sin \phi_{c'b} } = 0$ \cite{coeff}.

Consider squeezed states as input ($|\Psi^{\rm{in}} \rangle = [\hat{D}(\alpha)\hat{S}(r)|0\rangle]^{\otimes N}$), with the number of transmission channels $N$, displacement operator $\hat{D}(\alpha) = e^{\alpha \hat{a}^{\dagger} - \alpha^{\ast} \hat{a}}$, and squeezing operator $\hat{S}(r) = e^{ (r /2)(\hat{a}^{\dagger 2} - \hat{a}^{2}) }$ (complex number $\alpha$ being the amplitude and real number $r$ the squeezing parameter). One can obtain $\langle(\Delta \hat{x}_{a'}^{\rm{in}})^2 \rangle = e^{-2r}$, $\langle(\Delta \hat{x}_{b'}^{\rm{in}})^2 \rangle = \langle(\Delta \hat{p}_{b'}^{\rm{in}})^2 \rangle = \langle(\Delta \hat{x}_{c'}^{\rm{in}})^2 \rangle = \langle(\Delta \hat{p}_{c'}^{\rm{in}})^2 \rangle = 1$, ${\rm{cov}} (\hat{x}_{a'}^{\rm{in}}, \hat{x}_{a''}^{\rm{in}})|_{a' \neq a''} = 0$. Straightforwardly, Eq. (\ref{mean002az}) can be simplified to
\begin{eqnarray}
\label{sqz002az}
\overline {\langle (\Delta \hat{x}_b^{\rm{w}})^2 \rangle}_{\rm{sqz}}
= 2 \overline {V_{b} } + 1 -  \overline {T_{b}} (1-e^{-2r}),
\end{eqnarray}
since $\overline {T_{ b}} +\overline {R_{b}}-\overline{V_{b}} = 1$ (see Appendix A). It is obvious that with the increase of $r$, the modified average quantum fluctuation in Eq. (\ref{sqz002az}) decreases for a given RAM.

\begin{figure*}[tbh]
\begin{center}
\includegraphics[width=0.8\textwidth]{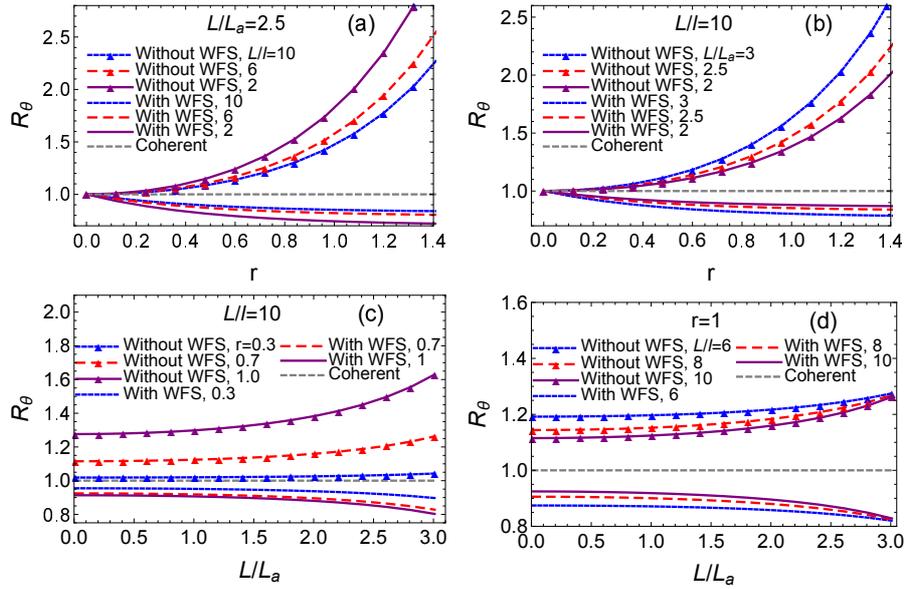} {}
\end{center}
\caption{The rescaled average quantum fluctuations $R_{\theta}$ ($\theta = \hat{x}_b$, $\hat{x}_b^{\rm{w}}$) versus $r$ [(a), (b)] and $L/L_a$ [(c), (d)]. The curves with the triangle marks denote the situation without WFS while the ones without the triangle marks (except for the dashed-gray line) represent the situation with WFS. The gray-dashed line stands for the average output noise with the coherent-state input. Parameters used are: (a) $L/L_a = 2.5$, (b) $L/l = 10$, (c) $L/l = 10$, and (d) $r = 1$.} 
\label{fig003}
\end{figure*}

Consider $r=0$ (i.e., the coherent-state input), Eq. (\ref{sqz002az}) is then reduced to $\overline {\langle (\Delta\hat{x}_b^{\rm{w}})^2 \rangle}_{\rm{sqz}\to\rm{\rm{coh}}} = 2  \overline {V_{b} } + 1$. For convenience, one can rewrite Eq. (\ref{sqz002az}) as
\begin{eqnarray}
\label{sqz003}
\overline {\langle (\Delta \hat{x}_b^{\rm{w}})^2 \rangle}_{\rm{sqz}}=  \overline {\langle (\Delta \hat{x}_b^{\rm{w}})^2 \rangle}_{\rm{coh}} -  \overline {T_{ b}} (1-e^{-2r}),
\end{eqnarray}
where $ \overline {\langle (\Delta \hat{x}_b^{\rm{w}})^2 \rangle}_{\rm{coh}} \equiv  2 \overline {V_{b} } + 1$. One can find that $\overline {\langle (\Delta\hat{x}_b^{\rm{w}})^2 \rangle}_{\rm{sqz}} < \overline {\langle (\Delta\hat{x}_b^{\rm{w}})^2 \rangle}_{\rm{coh}}$ always succeeds when $r>0$, which indicates that with WFS, the squeezed state has lower average output noise than that of the coherent state.

For comparison, we calculate the average quantum fluctuation in the absence of WFS
\begin{eqnarray}
\label{sqz001}
\overline {\langle (\Delta \hat{x}_b)^2 \rangle}_{\rm{sqz}} =  \overline {\langle (\Delta \hat{x}_b)^2 \rangle}_{\rm{coh}} +  \overline {T_{ b}} [\cosh (2r) - 1],
\end{eqnarray}
where $\overline {\langle (\Delta \hat{x}_b)^2 \rangle}_{\rm{coh}} \equiv 2  \overline {V_{ b} } + 1$ represents the average quantum fluctuation of the scattered light in the absence of WFS with the coherent-state input. The detailed derivation is present in Appendix B. 

Comparing Eqs. (\ref{sqz003}) and (\ref{sqz001}), one can extract the difference between the average quantum fluctuations with and without WFS
\begin{eqnarray}
\label{sqz004}
\overline {\langle (\Delta \hat{x}_b)^2 \rangle}_{\rm{sqz}} - \overline {\langle (\Delta \hat{x}_b^{\rm{w}})^2 \rangle}_{\rm{sqz}} =   \overline {T_{ b}} \sinh (2r),
\end{eqnarray}
since $\overline {\langle (\Delta \hat{x}_b)^2 \rangle}_{\rm{coh}} = \overline {\langle (\Delta\hat{x}_b^{\rm{w}})^2 \rangle}_{\rm{coh}}$ is used. Fig. \ref{fig002}(a) [\ref{fig002}(b)] depicts the difference between $\overline{\langle (\Delta \hat{x}_b)^2 \rangle}$ and $\overline{\langle (\Delta \hat{x}_b^{\rm{w}})^2 \rangle}$ as a function of $L/L_a$ and $r$ [$L/L_a$ and $L/l$]. It is found that the difference is always larger than zero for a squeezed-state input ($r>0$) which implies that the WFS can reduce the average quantum fluctuation in the presence of a squeezed-state input.

For convenience, we introduce the rescaled average quantum fluctuation, 
\begin{eqnarray}
R_{\theta} = \overline{\langle (\Delta \theta )^2 \rangle}_{\rm{sqz}} / \overline{\langle (\Delta \theta )^2 \rangle}_{\rm{coh}},
\end{eqnarray}
where $\theta = \hat{x}_b, \hat{x}_b^{\rm{w}}$. Fig. \ref{fig003} compares the rescaled average quantum fluctuations $R_{\theta}$ with and without WFS as a function of $r$ [(a), (b)] and $L/L_a$ [(c), (d)]. In Figs. \ref{fig003}(a)-\ref{fig003}(d), the curves with the triangle marks denote the situations without WFS whereas those without the triangle marks (except for the gray-dashed line) represent the cases with WFS. The gray-dashed line stands for the average output noise for the coherent-state input.

As shown in Figs. \ref{fig003}(a)-\ref{fig003}(d), the purple-solid, red-dashed, and blue-dashed lines without triangle marks are always below their corresponding lines with triangle marks when $r>0$. This implies that the WFS can always reduce the average quantum noise for the squeezed-state input.

\section{Comparison and discussion}

\begin{figure*}[tbh]
\begin{center}
\includegraphics[width=0.8\textwidth]{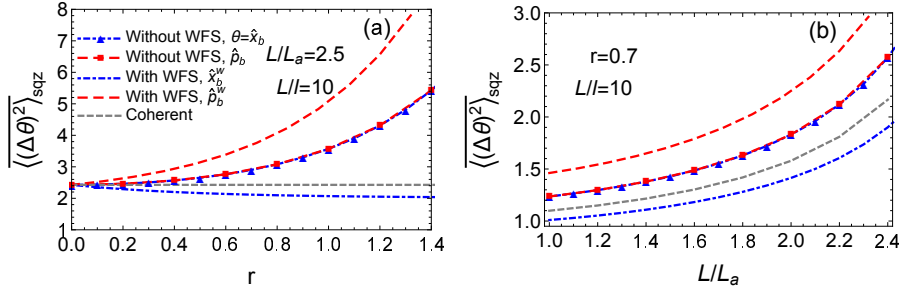} {}
\end{center}
\caption{The average quantum fluctuations $\overline{\langle (\Delta \theta)^2 \rangle} $ ($\theta = \hat{x}_b$, $\hat{p}_b$, $\hat{x}_b^{\rm{w}}$, or $\hat{p}_b^{\rm{w}}$) versus (a) $r$ and (b) $L/L_a$. The curves with the triangle (square) marks denote the situation of $\theta = \hat{x}_b$ ($\theta = \hat{p}_b$) while the blue-dashed-dotted (red-dashed) line represents the situation of $\theta = \hat{x}_b^{\rm{w}}$ ($\theta = \hat{p}_b^{\rm{w}}$). The gray-dashed line labeled ``coherent'' stands for the average output noise for the coherent-state input. Parameters used are: (a) $L/L_a = 2.5$, $L/l=10$, and (b) $r=0.7$, $L/l = 10$.} 
\label{fig003a}
\end{figure*}

\subsection{The suppressed and increased average quantum fluctuations}

One important aspect neglected so far is the quadrature $\hat{p}_{b}^{\rm{w}}$. Similar to $\hat{x}_{b}^{\rm{w}}$, the average quantum fluctuation of $\hat{p}_{b}^{\rm{w}}$ can be cast into
\begin{eqnarray}
\overline {\langle (\Delta \hat{p}_b^{\rm{w}})^2 \rangle}_{\rm{sqz}}
= \overline {\langle (\Delta \hat{p}_b^{\rm{w}})^2 \rangle}_{\rm{coh}} + \overline {T_{ b}} (e^{2r} - 1),
\label{sqz002pp}
\end{eqnarray}
where $\overline {\langle (\Delta \hat{p}_b^{\rm{w}})^2 \rangle}_{\rm{coh}}= 2  \overline {V_{ b} } + 1$ means the average output quantum fluctuation for the coherent-state input. Meanwhile, the average quantum fluctuation of $\hat{p}_b$ in the absence of WFS is found to be
\begin{eqnarray}
\label{sqz001aa}
\overline {\langle (\Delta \hat{p}_b)^2 \rangle}_{\rm{sqz}} =  \overline {\langle (\Delta \hat{p}_b)^2 \rangle}_{\rm{coh}} +  \overline {T_{ b}} [\cosh (2r) - 1],
\end{eqnarray}
where $\overline {\langle (\Delta \hat{p}_b)^2 \rangle}_{\rm{coh}} = 2  \overline {V_{ b} } + 1$ represents the case of the coherent-state input.

Figs. \ref{fig003a}(a) and \ref{fig003a}(b) compare the average quantum fluctuations between $\hat{x}$ and $\hat{p}$ with and without WFS. Fig. \ref{fig003a}(a) plots the average quantum fluctuations as a function of $r$. The blue-dotted-dashed line with (without) triangle marks represents the case of $\hat{x}_b$ ($\hat{x}_b^{\rm{w}}$) while the red-dashed line with (without) square marks denotes the case of $\hat{p}_b$ ($\hat{p}_b^{\rm{w}}$). The gray-dashed line stands for the average quantum noise for the coherent-state input. It is easy to find that with the increase of $r$, $\overline{\langle (\Delta \hat{x}^{\rm{w}}_b)^2\rangle}$ decreases whereas $\overline{\langle (\Delta \hat{x}_b)^2\rangle}$, $\overline{\langle (\Delta \hat{p}_b)^2\rangle}$, and $\overline{\langle (\Delta \hat{p}^{\rm{w}}_b)^2\rangle}$ increase. In the absence of WFS, the average quantum fluctuations of $ \hat{x}_b$ and $ \hat{p}_b$ coincide with each other. Intriguingly, in the presence of WFS, the average quantum fluctuation of $ \hat{x}_b^{\rm{w}}$ is smaller than that of $ \hat{x}_b$ whereas the average quantum fluctuation of $ \hat{p}_b^{\rm{w}}$ becomes larger than that of $ \hat{p}_b$, which yields that the WFS leads to a decrease in the average quantum fluctuation of $\hat{x}$ but an increase in that of $\hat{p}$. In the absence of WFS, the squeezed light experiences random phases when propagating through the RAM. This means that both $\hat{x}_b$ and $\hat{p}_b$ will be the mixture of the squeezing, anti-squeezing, and quadrature component in various orientation of the original squeezed light, which leads to the same fluctuation for both quadrature. On the other hand, the WFS removes these random phases, which results in $\hat{x}_b^{\rm{w}}$ and $\hat{p}_b^{\rm{w}}$ retaining the original quadrature of the squeezed light, with additional noise added from the spontaneous emission.


Fig. \ref{fig003a}(b) depicts the average quantum fluctuations as a function of $L/L_a$. It can be seen that as $L/L_a$ increases, $\overline{\langle (\Delta \hat{x}_b)^2\rangle}$, $\overline{\langle (\Delta \hat{x}^{\rm{w}}_b)^2\rangle}$, $\overline{\langle (\Delta \hat{p}_b)^2\rangle}$, and $\overline{\langle (\Delta \hat{p}^{\rm{w}}_b)^2\rangle}$ increase. Notably, $\overline{\langle (\Delta \hat{x}_b^{\rm{w}})^2 \rangle}$ is still below the gray-dashed line, which indicates that the squeezed state has lower average output quantum noise than that of the coherent state with WFS.

\begin{figure*}[tbh]
\begin{center}
\includegraphics[width=0.8\textwidth]{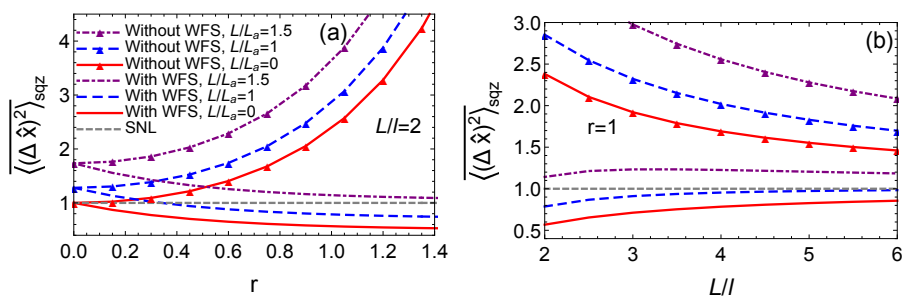} {}
\end{center}
\caption{The average output quantum fluctuations $\overline{\langle (\Delta \hat{x})^2 \rangle}_{\rm{sqz}}$ ($\hat{x} = \hat{x}_b, \hat{x}_b^{\rm{w}} $) versus (a) $r$ and (b) $L/l$. The blue triangle-marked dashed, blue dashed, red triangle-marked solid, red solid, and gray dashed-dotted curves denote the cases of amplifying media without WFS, amplifying ones with WFS, linear ones without WFS, linear ones with WFS, and the shot noise, respectively. Parameters used are: (a) $L/L_a = 1,$ $L/l = 2$ for amplifying media and $L/L_a = 0,$ $L/l = 2$ for linear ones, (b) $L/L_a = 1,$ $r = 1$ for amplifying ones and $L/L_a = 0,$ $r = 1$ for linear ones.}
\label{xr}
\end{figure*}

\subsection{Comparison between the amplifying and linear cases}

To give insight into the effects of nonlinearity on the suppressed quantum noise, we compare the amplifying and linear situations. By setting $L_a \to \infty$ (i.e., $ L/L_a \to 0$ and $\overline{V_{b}} = 0 $, amplifying effects vanishing), based on Eqs. (\ref{sqz002az}) and (\ref{sqz001}), the average quantum fluctuations in linear cases with and without WFS can be expressed as
\begin{eqnarray}
\label{linear00a}
\overline {\langle (\Delta \hat{x}_b^{\rm{w}})^2 \rangle}_{\rm{sqz},\rm{lin}}=&1 - \overline {T_{ b}} (1-e^{-2r}), \\
\overline {\langle (\Delta\hat{x}_b)^2 \rangle}_{\rm{sqz},\rm{lin}} =& 1 +  \overline {T_{ b}} [\cosh (2r) - 1],
\end{eqnarray}
respectively, which is consistent with our previous work \cite{li2019b}.

Fig. \ref{xr} shows the average output quantum fluctuations $\overline{\langle (\Delta \hat{x})^2 \rangle}_{\rm{sqz}}$ ($\hat{x} = \hat{x}_b, \hat{x}_b^{\rm{w}}$) versus (a) $r$ and (b) $L/l$. The blue-triangle-marked-dashed, blue-dashed, red-triangle-marked-solid, red-solid, and gray-dotted-dashed curves denote the cases of amplifying media without WFS, amplifying ones with WFS, linear ones without WFS, linear ones with WFS, and the SNL, respectively. 

In Fig. \ref{xr}(a), with the increasing of $r$, the average output quantum noise without WFS increases whereas the one with WFS decreases. This is due to the fact that the average output quantum noise without WFS is related to the input quantum noise, which encompasses not only the noise of squeezed quadrature ($\langle (\Delta \hat{x}_{a'}^{\rm{in}})^2 \rangle = e^{-2r}$) but also the noise of anti-squeezed quadrature ($\langle (\Delta \hat{p}_{a'}^{\rm{in}})^2 \rangle = e^{2r}$). When $r$ becomes large, the maximum noise ascends steeply, which provokes the increase of the average output quantum noise. By contrast, the average output quantum noise with WFS is related to the input quantum noise, which includes only the squeezed noise ($\langle (\Delta \hat{x}_{a'}^{\rm{in}})^2 \rangle = e^{-2r}$). The noise of the anti-squeezed quadrature $\langle (\Delta \hat{p}_{a'}^{\rm{in}})^2 \rangle = e^{2r}$ disappears owning to the destructive interference of quantum noise \cite{li2019b,elste2009}. With the increase of $r$, the squeezed noise ($\langle (\Delta \hat{x}_{a'}^{\rm{in}})^2 \rangle = e^{-2r}$) decreases, which gives rise to a decrease in the average output quantum noise. Fig. \ref{xr}(b) shows the average output noise as a function of $L/l$. It is obvious that, with the increase of $L/l$, the average output noise without WFS decreases, whereas the average output noise with WFS increases.

From Figs. \ref{xr}(a) and \ref{xr}(b), it is found that WFS can reduce the average quantum fluctuation in both linear and amplifying cases. Nevertheless, unlike the linear case ($L/l = 0$, solid lines) where the suppressed quantum noise can always reach below the SNL, the reduced average quantum noise can be either below or above the SNL for the amplifying case. 

The phenomenon can be explained as follows: initially, the linear situation without WFS shows the excess noise resulting from multiple scattering. On the contrary, the amplifying case without WFS presents the excess noise induced by not only multiple scattering but also nonlinear amplification. The WFS can effectively reduce the excess noise from multiple scattering rather than amplification. Resultantly, for the linear case, the excess noise is well suppressed below the SNL via WFS. Nevertheless, for the amplifying case, the excess noise can be reduced below the SNL only if the multiple scattering dominates (i.e., weak amplification strength). It is worth noting that the excess noise from amplification can be categorized into two types: one from spontaneous emission and the other one from stimulated emission. Although the WFS is not able to reduce the noise due to spontaneous emission, it can still effectively suppress the one from stimulated emission. Therefore, the WFS could still reduce the output average noise below the SNL in some certain condition for the amplifying case.

\subsection{The condition for the suppressed quantum fluctuations to achieve below the SNL}

\begin{figure}[tbh]
\begin{center}
\includegraphics[width=.40\textwidth]{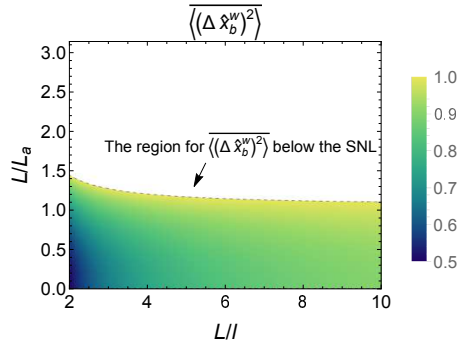} {}
\end{center}
\caption{The region for the suppressed average quantum fluctuation to reach below the SNL. Parameter: $e^{-2r} \to 0$. }
\label{figsnl}
\end{figure}

Sub-shot noise of light belongs to the most prominent nonclassical trait. Nevertheless, the suppressed average quantum noise can not always reach below the SNL for the amplifying case. We shall now discuss the condition for the suppressed average quantum noise to achieve below the SNL. Assuming that the suppressed average quantum noise reaches below the SNL (i.e., $\overline {\langle (\Delta \hat{x}_b^{\rm{w}})^2 \rangle}_{\rm{sqz}} < 1$), from Eq. (\ref{sqz002az}), one has 
\begin{eqnarray}
\label{snl002}
2  \overline {V_{ b} } -\overline {T_{ b}} (1-e^{-2r}) < 0.
\end{eqnarray}

Combined with Eqs. (\ref{eq2a}) and (\ref{eq2c}), Eq. (\ref{snl002}) can be rewritten as
\begin{eqnarray}
\sin (\frac{l}{L_a}) (1+e^{-2r}) + 2 \sin (\frac{L-l}{L_a}) - 2 \sin (\frac{L}{L_a}) <0.
\end{eqnarray}
The solution is found to be
\begin{eqnarray}
\label{solution001}
	L/L_a < \arcsin M_{+},
\end{eqnarray}
where the detailed derivation is shown in Appendix C, $M_{+} = [mn + \sqrt{4m^2-n^2+4}]/[2(m^2 +1)]$ with $n = 1+e^{-2r}$, $m = (1 - \sqrt{1 - p^2})/p$ and $p = \sin (l/L_a)$. Fig. \ref{figsnl} intuitively illustrates the solution in Eq. (\ref{solution001}) where for simplicity we consider the situation of the large squeezing strength ($e^{-2r} \to 0$) \cite{sqzpara}. The colored region allows $\overline{ \langle (\Delta \hat{x}_b^{\rm{w}})^2 \rangle }_{\rm{sqz}}$ to reach below the SNL. It is obvious that this condition requires a weak amplification strength. This is because the WFS can effectively reduce the excess noise induced by multiple scattering but not amplification.

\section{Conclusion}

In summary, we investigate the effect of wavefront shaping on the average quantum noise of scattered modes in the random amplifying media. It is demonstrated that the wavefront shaping offers the ability to reduce the average output quantum noise for a squeezed-state input. Particularly, the wavefront shaping can effectively suppress the excess noise resulted from multiple scattering but not amplification. This reduction is owing to the destructive interference of quantum noise. In addition, both the decrease on amplification strength \textit{and} the increase on the input squeezing strength can lead to a decrease in the suppressed average noise. 

It is found that unlike the linear media where the suppressed average quantum noise is always below the shot-noise level, the reduced average quantum noise can be either below or above the shot-noise level for the amplifying case. Moreover, we provide the condition for the suppressed noise to achieve below the shot-noise level which requires the amplification strength to be weak. Our results may have potential implications in quantum information processing, such as high-resolution imaging and optical authentication. For instance, in the authentication system based on scattering medium \cite{goorden2014,yao2016}, the most vital process involves light focusing through a random medium with the help of WFS. Our work might contribute to design this kind of authentication schemes with squeezed-state input.

\appendix
\section{Derivation of the summation of transmission, reflection, and spontaneous emission coefficients}
\label{derivation}
The input-output relation of a random amplifying medium is given by
\begin{eqnarray}
\hat{a}_b^{\dagger} = \sum_{a'} t_{a' b}^{\ast} \hat{a}_{a'}^{\rm{in} \dagger} + \sum_{b'} r_{b' b}^{\ast} \hat{a}_b^{\rm{in} \dagger} + \sum_{c'} v_{c' b}^{\ast} \hat{c}_{c'}^{\rm{in} }, \\ \nonumber
\hat{a}_b = \sum_{a'} t_{a' b} \hat{a}_{a'}^{\rm{in}} + \sum_{b'} r_{b' b} \hat{a}_{b'}^{\rm{in}} + \sum_{c'} v_{c' b}^{\ast} \hat{c}_{c'}^{{\rm{in}}\dagger},
\end{eqnarray}
where $t_{a'b}^{\ast}$ ($r_{b'b}^{\ast}$, $v_{c' b}^{\ast}$) is the conjugate of $t_{a'b}$ ($r_{b'b}$, $v_{c' b}$). According to the commutation relation $[\hat{a}_b, \hat{a}_b^{\dagger}] = 1$, one can easily obtain 
\begin{eqnarray}
\label{sumsumsum}
\sum_{a'} |t_{a'b}|^2 + \sum_{b'} |r_{b'b}|^2 - \sum_{c'} |v_{c'b}|^2 = 1,
\end{eqnarray}
where $[\hat{a}_{i}^{\rm{in}}, \hat{a}_{j}^{\rm{in}\dagger}] = \delta_{ij}$ ($i,j=a',b'$) has been used. Let $T_{b} = \sum_{a'}|t_{a'b}|^2$, $R_{b} = \sum_{b'}|r_{b'b}|^2$, and $V_{b} = \sum_{c'} |v_{c'b}|^2$. Eq. (\ref{sumsumsum}) can be then rewritten as $ T_{b} +  R_{b} - V_{b} = 1$.

\section{Average quantum fluctuation of the scattered light in the absence of WFS}

\label{cohinput}
The variance of $\hat{x}_b$ without WFS is given by
\begin{eqnarray}
\label{v2b}
\langle (\Delta \hat{x}_b)^2\rangle = \langle \hat{x}_b^2\rangle - \langle \hat{x}_b\rangle^2.
\end{eqnarray}
To obtain the variance $\langle (\Delta \hat{x}_b)^2\rangle$, it is necessary to calculate $ \langle \hat{x}_b\rangle$ and $\langle \hat{x}_b^2\rangle$.

In the absence of WFS, the mean value of $\hat{x}_b$ in Eq. (\ref{xb}) is found to be
\begin{eqnarray}
\langle \hat{x}_b \rangle =& \sum_{a'} {\sqrt{T_{a'b}} [ \cos \phi_{a'b} \langle \hat{x}_{a'}^{\rm{in}} \rangle - \sin \phi_{a'b} \langle \hat{p}_{a'}^{\rm{in}} \rangle] },
\end{eqnarray}

According to Eq. (\ref{xb}), $\hat{x}_b^2$ can be obtained
\begin{eqnarray}
\label{xb0}
\hat{x}_b^2 &=& \sum_{a'a''} \sqrt{T_{a' b}T_{a'' b}} [\cos \phi_{a' b} \cos \phi_{a'' b} \hat{x}_{a'}^{\rm{in}} \hat{x}_{a''}^{\rm{in}} \nonumber \\
&&+ \sin \phi_{a' b} \sin \phi_{a'' b} \hat{p}_{a'}^{\rm{in}} \hat{p}_{a''}^{\rm{in}}  \nonumber \\
&&- \cos \phi_{a' b} \sin \phi_{a'' b} \hat{x}_{a'}^{\rm{in}} \hat{p}_{a''}^{\rm{in}} \nonumber \\
&&-\sin \phi_{a' b}\cos \phi_{a'' b} \hat{p}_{a'}^{\rm{in}} \hat{x}_{a''}^{\rm{in}}] \nonumber \\
&&+\sum_{b'b''} \sqrt{R_{b' b} R_{b'' b}} [\cos \phi_{b' b}\cos \phi_{b'' b} \hat{x}_{b'}^{\rm{in}} \hat{x}_{b''}^{\rm{in}} \nonumber \\
&&+ \sin \phi_{b' b} \sin \phi_{b'' b} \hat{p}_{b'}^{\rm{in}} \hat{p}_{b''}^{\rm{in}} \nonumber \\
&&- \cos \phi_{b' b}  \sin \phi_{b'' b} \hat{x}_{b'}^{\rm{in}} \hat{p}_{b''}^{\rm{in}} \nonumber \\
&&-   \sin \phi_{b' b}  \cos \phi_{b'' b} \hat{p}_{b'}^{\rm{in}} \hat{x}_{b''}^{\rm{in}}] \nonumber \\
&&+  \sum_{c'c''} \sqrt{V_{c' b}V_{c'' b}} [\cos \phi_{c' b} \cos \phi_{c'' b} \hat{x}_{c'}^{\rm{in}} \hat{x}_{c''}^{\rm{in}} \nonumber \\
&&+ \sin \phi_{c' b} \sin \phi_{c'' b} \hat{p}_{c'}^{\rm{in}} \hat{p}_{c''}^{\rm{in}} \nonumber \\
&&- \cos \phi_{c' b} \sin \phi_{c'' b} \hat{x}_{c'}^{\rm{in}}  \hat{p}_{c''}^{\rm{in}} \nonumber \\
&&-  \sin \phi_{c' b} \cos \phi_{c'' b}  \hat{p}_{c'}^{\rm{in}} \hat{x}_{c''}^{\rm{in}} ]\nonumber \\
&&+ \sum_{a'b'} 2\sqrt{T_{a' b}R_{b' b}} \{[\cos \phi_{a' b} \hat{x}_{a'}^{\rm{in}} - \sin \phi_{a' b} \hat{p}_{a'}^{\rm{in}}]\nonumber \\
&& \times[\cos \phi_{b' b} \hat{x}_{b'}^{\rm{in}} - \sin \phi_{b' b} \hat{p}_{b'}^{\rm{in}}]\} \nonumber \\
&&+ \sum_{a'c'} 2\sqrt{T_{a' b}V_{c' b}} \{[\cos \phi_{a' b} \hat{x}_{a'}^{\rm{in}} - \sin \phi_{a' b} \hat{p}_{a'}^{\rm{in}}]\nonumber \\
&& \times[\cos \phi_{c' b} \hat{x}_{c'}^{\rm{in}} - \sin \phi_{c' b} \hat{p}_{c'}^{\rm{in}}]\} \nonumber \\
&&+ \sum_{b'c'} 2\sqrt{R_{b' b}V_{c' b}}\{[\cos \phi_{b' b} \hat{x}_{b'}^{\rm{in}} - \sin \phi_{b' b} \hat{p}_{b'}^{\rm{in}}]\nonumber \\
&&\times[\cos \phi_{c' b} \hat{x}_{c'}^{\rm{in}} - \sin \phi_{c' b} \hat{p}_{c'}^{\rm{in}}]\}.
\end{eqnarray}

Then $\langle \hat{x}_b^2 \rangle$ is found to be
\begin{eqnarray}
\langle \hat{x}_b^2 \rangle &= & \sum_{a'a''} \sqrt{T_{a' b}T_{a'' b}} [\cos \phi_{a' b} \cos \phi_{a'' b} \langle\hat{x}_{a'}^{\rm{in}} \hat{x}_{a''}^{\rm{in}}\rangle \nonumber \\
&&+ \sin \phi_{a' b} \sin \phi_{a'' b} \langle\hat{p}_{a'}^{\rm{in}} \hat{p}_{a''}^{\rm{in}}\rangle \nonumber \\
&&- \cos \phi_{a' b} \sin \phi_{a'' b} \langle\hat{x}_{a'}^{\rm{in}} \hat{p}_{a''}^{\rm{in}} \rangle \nonumber \\
&&-\sin \phi_{a' b}\cos \phi_{a'' b} \langle\hat{p}_{a'}^{\rm{in}} \hat{x}_{a''}^{\rm{in}} \rangle] \nonumber \\
&&+\sum_{b'} R_{b' b}  [\cos^2 \phi_{b' b} \langle \hat{x}_{b'}^{\rm{in} 2} \rangle + \sin^2 \phi_{b' b} \langle \hat{p}_{b'}^{\rm{in} 2} \rangle \nonumber \\
&&- \cos \phi_{b' b}  \sin \phi_{b' b} \langle \hat{x}_{b'}^{\rm{in}} \hat{p}_{b'}^{\rm{in}} + \hat{p}_{b'}^{\rm{in}} \hat{x}_{b'}^{\rm{in}} \rangle] \nonumber \\
&&+  \sum_{c'} V_{c' b} [\cos^2 \phi_{c' b} \langle \hat{x}_{c'}^{\rm{in} 2} \rangle + \sin^2 \phi_{c' b} \langle  \hat{p}_{c'}^{\rm{in} 2} \rangle  \nonumber \\
&&- \cos \phi_{c' b} \sin \phi_{c' b} \langle \hat{x}_{c'}^{\rm{in}}  \hat{p}_{c'}^{\rm{in}} +  \hat{p}_{c'}^{\rm{in}} \hat{x}_{c'}^{\rm{in}}\rangle ].
\label{xb2}
\end{eqnarray}

The variance is then expressed as
\begin{eqnarray}
\langle (\Delta \hat{x}_{b})^2 \rangle &=& 
\sum_{a'}T_{a'b}[\cos^2 \phi_{a'b}\langle (\Delta \hat{x}_{a'}^{\rm{in}})^2 \rangle \nonumber \\
&&+ \sin^2 \phi_{a'b}\langle (\Delta \hat{p}_{a'}^{\rm{in}})^2 \rangle \nonumber \\
&&- 2\cos \phi_{a'b} \sin \phi_{a'b} {\rm{cov}}(\hat{x}_{a'}^{\rm{in}}, \hat{p}_{a'}^{\rm{in}}))]\nonumber \\
&&   + \sum_{b'}R_{b'b}[\cos^2 \phi_{b'b}\langle (\Delta \hat{x}_{b'}^{\rm{in}})^2 \rangle \nonumber \\
&&+ \sin^2 \phi_{b'b}\langle (\Delta \hat{p}_{b'}^{\rm{in}})^2 \rangle \nonumber \\
&&- 2\cos \phi_{b'b} \sin \phi_{b'b} {\rm{cov}}(\hat{x}_{b'}^{\rm{in}}, \hat{p}_{b'}^{\rm{in}}) ]  \nonumber \\
&&+ \sum_{c'}V_{c'b}[\cos^2 \phi_{c'b}\langle (\Delta \hat{x}_{c'}^{\rm{in}})^2 \rangle\nonumber \\
&& + \sin^2 \phi_{c'b}\langle (\Delta \hat{p}_{c'}^{\rm{in}})^2 \rangle\nonumber \\
&& - 2\cos \phi_{c'b} \sin \phi_{c'b} {\rm{cov}}(\hat{x}_{c'}^{\rm{in}}, \hat{p}_{c'}^{\rm{in}}) ] \nonumber \\
&&+ \sum_{a'\neq a''} \{ \sqrt{T_{a'b} T_{a''b}} [2\cos \phi_{a' b} \cos \phi_{a''b}  \nonumber \\
&&\times{\rm{cov}} (\hat{x}_{a'}^{\rm{in}}, \hat{x}_{a''}^{\rm{in}}) + 2\sin \phi_{a' b} \sin \phi_{a''b}  {\rm{cov}}(\hat{p}_{a'}^{\rm{in}}, \hat{x}_{a''}^{\rm{in}})  \nonumber \\
&&-2\cos \phi_{a'b} \sin \phi_{a''b} {\rm{cov}}(\hat{x}_{a'}^{\rm{in}}, \hat{p}_{a''}^{\rm{in}})] \},
\label{xb2a}
\end{eqnarray}
where the covariance fuction is defined as ${\rm{cov}}(\hat{Y},\hat{Z}) \equiv \frac{1}{2} (\langle \hat{Y} \hat{Z} \rangle +\langle \hat{Z} \hat{Y} \rangle) - \langle \hat{Y}\rangle \langle \hat{Z} \rangle $.

Consider the squeezed state as input ($|\Psi^{\rm{in}} \rangle = [\hat{D}(\alpha)\hat{S}(r)|0\rangle]^{\otimes N}$), by averaging over realizations of disorder, Eq. (\ref{xb2a}) can be simplified as
\begin{eqnarray}
\overline{\langle(\Delta\hat{x}_b)^2\rangle}_{\rm{sqz}}&=&\frac{1}{2}\overline {T_{ b}} [\overline{\langle (\Delta\hat{x}_{a'}^{\rm{in} })^2 \rangle }+\overline{\langle (\Delta \hat{p}_{a'}^{\rm{in} })^2\rangle }] +\overline {R_{b}} +\overline {V_{ b} } \nonumber \\
&= & 2\overline {V_{ b} } + 1 +\overline{T_{b}} [\cosh (2r) -1].
\label{sqz001aaa}
\end{eqnarray}
When the input is the coherent state (i.e., $r=0$), Eq. (\ref{sqz001aaa}) can be cast into
\begin{eqnarray}
\overline {\langle (\Delta\hat{x}_b)^2 \rangle}_{\rm{coh}} =  2  \overline {V_{ b} } + 1 .
\label{coh001aaa}
\end{eqnarray}

\section{Derivation of Eq. (\ref{solution001})}
\label{appendixc}

Inserting Eqs. (\ref{eq2a}) and (\ref{eq2c}) into (\ref{snl002}) arrives at
\begin{eqnarray}
\label{x01}
\sin (\frac{l}{L_a}) (1 +e^{-2 r}) + 2 \sin (\frac{L-l}{L_a}) - 2 \sin (\frac{L}{L_a})<0.
\end{eqnarray}
By using trigonometric formulas, Eq. (\ref{x01}) could be expressed as 
\begin{eqnarray}
\label{x03}
[1 + e^{-2r} - 2\cos(\frac{L}{L_a})] \sin (\frac{l}{L_a}) + [\cos (\frac{l}{L_a}) - 1] 2 \sin (\frac{L}{L_a}) <0.
\end{eqnarray}
Eq. (\ref{x03}) could be further cast into
\begin{eqnarray}
\label{x05}
\frac{1+e^{-2r}}{2} -  \frac{1- \sqrt{1-p^2}}{p} M <  \sqrt{1 - M^2},
\end{eqnarray}
where we have defined $M = \sin(L/L_a)$ and $p = \sin(l/L_a)$ for simplicity.

From Eq. (\ref{x05}), one can obtain
\begin{eqnarray}
\label{x08}
(m^2+1) M^2 -  m n  M - (4 - n^2)/4 < 0,
\end{eqnarray}
where $n = 1 + e^{-2r}$ and $m = (1- \sqrt{1-p^2})/p$.

It is easy to verify that there always exists a solution for Eq. (\ref{x08})
\begin{eqnarray}
M_{-} < M < M_{+},
\end{eqnarray}
where 
\begin{eqnarray}
M_{\pm} = \frac{m n \pm \sqrt{ 4m^2 - n^2 + 4}}{2 (m^2 + 1)}.
\end{eqnarray}
It is worthy pointing out that $M_{-}<0$ and $M_{+}>0$. However, in our scheme, $M= \sin (L/L_a)>0$ should be positive (since $L/L_a < \pi$ corresponding to the case below the laser threshold \cite{fedorov2009}). Therefore, the solution is found to be
\begin{eqnarray}
0< M < M_{+}.
\end{eqnarray}
Accordingly, one can obtain
\begin{eqnarray}
L/L_a < \arcsin(M_{+}).
\end{eqnarray}



\end{document}